\documentclass[conference]{IEEEtran}
\IEEEoverridecommandlockouts
\usepackage{cite}
\usepackage{amsmath,amssymb,amsfonts}
\usepackage{algorithmic}
\usepackage{graphicx}
\usepackage{textcomp}
\usepackage{xcolor}
\usepackage{subcaption}
\usepackage{tikz}
\usepackage{hyperref}
\def\BibTeX{{\rm B\kern-.05em{\sc i\kern-.025em b}\kern-.08em
    T\kern-.1667em\lower.7ex\hbox{E}\kern-.125emX}}
\def\VSPACE{0mm}

\newcommand\copyrighttext{
	\footnotesize \textcopyright 2021 IEEE. Personal use of this material is permitted.
	Permission from IEEE must be obtained for all other uses, in any current or future 
	media, including reprinting/republishing this material for advertising or promotional 
	purposes, creating new collective works, for resale or redistribution to servers or 
	lists, or reuse of any copyrighted component of this work in other works. 
	DOI: \href{https://doi.org/10.1109/ISCAS51556.2021.9401621}{10.1109/ISCAS51556.2021.9401621} }
\newcommand\copyrightnoticeOwn{%
	\begin{tikzpicture}[remember picture,overlay]
		\node[anchor=north,yshift=-10pt] at (current page.north) {\fbox{\parbox{\dimexpr\textwidth-\fboxsep-\fboxrule\relax}{\copyrighttext}}};
	\end{tikzpicture}%
}

\begin{document}

\title{Robust Deep Neural Object Detection and Segmentation for Automotive Driving Scenario with Compressed Image Data\\
\thanks{The authors gratefully acknowledge that this work has been supported by the Deutsche Forschungsgemeinschaft (DFG) under contract number KA 926/10-1.}
}

\author{\IEEEauthorblockN{Kristian Fischer, Christian Blum, Christian Herglotz, and Andr\'e Kaup\\}
	\IEEEauthorblockA{\textit{Multimedia Communications and Signal Processing} \\
		\textit{Friedrich-Alexander-Universit\"at Erlangen-N\"urnberg (FAU)}\\
		Cauerstr. 7, 91058 Erlangen, Germany \\
		\{Kristian.Fischer, Christian.Blum, Christian.Herglotz, Andre.Kaup\}@fau.de}
}
\maketitle
\copyrightnoticeOwn

\begin{abstract}
Deep neural object detection or segmentation networks are commonly trained with pristine, uncompressed data. However, in practical applications the input images are usually deteriorated by compression that is applied to efficiently transmit the data. Thus, we propose to add deteriorated images to the training process in order to increase the robustness of the two state-of-the-art networks Faster and Mask R-CNN. Throughout our paper, we investigate an autonomous driving scenario by evaluating the newly trained models on the \textit{Cityscapes} dataset that has been compressed with the upcoming video coding standard Versatile Video Coding (VVC). When employing the models that have been trained with the proposed method, the weighted average precision of the R-CNNs can be increased by up to 3.68 percentage points for compressed input images, which corresponds to bitrate savings of nearly 48\,~\%.
\end{abstract}


\section{Introduction}
\label{sec:intro}

With the beginning growth of autonomous driving systems, more and more scenarios arise where multiple cars are connected with an edge computing platform~\cite{liu2019, yuan2018}. There, the incoming sensor data is collected and intelligent algorithms are applied to predict e.g. possible hazardous situations from this data. With these predictions, the connected cars can be warned and counteractions can be initiated to reduce the risk of accidents. This data might include but is not limited to the velocity and position of the single cars, RADAR information, or video data from the cameras capturing the car surrounding. But even with the advent of 5G, the data transmitted from dozens of cars to a central computing platform point is too large to be transmitted as raw data. Therefore, suitable data compression is required. 

\begin{figure}[t]
	\centering
	\includegraphics[width=0.45\textwidth]{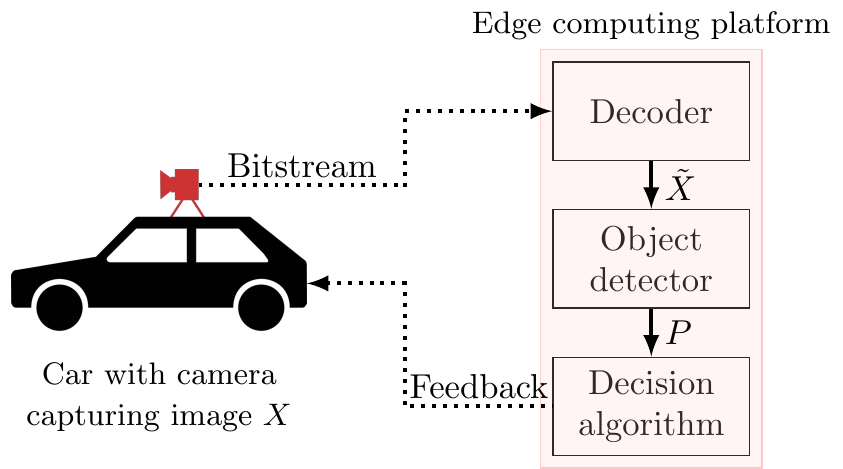}
	\caption{Edge computing scenario for one car transmitting an encoded image $X$ to an edge computing platform.}
	\label{fig:edge computing scenario}
		\vspace{-2mm}
\end{figure}

For the scope of our paper we focus on video data that is captured by the cameras inside and around a car. From this data, important information can be derived like other nearby road users. For that task, deep neural networks have made tremendous progress throughout the recent years detecting and classifying relevant objects $P$ from a single image~$X$. Two examples for this are the state-of-the-art object detection network Faster Region-based Convolutional Neural Network~(R-CNN)~\cite{ren2017} and its successor Mask R-CNN~\cite{he2017}, which additionally segments the objects pixel-wisely from the background. In the scenario presented in Fig.~\ref{fig:edge computing scenario}, the R-CNNs have to work with input images $\tilde{X}$ that are deteriorated by the compression that has been applied for an efficient transmission under bitrate constraints. Ultimately, the predicted objects $P$ can be fed into a decision algorithm together with other data to obtain valuable information about the events occurring on the streets, which can be shared with all connected cars via a feedback connection.

However, during the classic training process of the object detection networks, their models are trained with pristine data. Thus, to the best of our knowledge, this paper for the first time investigates whether the R-CNN models can be made more robust against compression artifacts by optimizing the training process with specific data augmentation or fine-tuning. As compression framework we selected the upcoming video coding standard Versatile Video Coding (VVC)~\cite{chen2020vtm10}.

The drawn scenario can be counted to the field of Video Coding for Machines (VCM), which aims for an efficient compression when the data, in this case image data, is directly analyzed by an algorithm rather than being observed by a human. To tackle this problem, MPEG founded an ad-hoc group in 2019~\cite{zhang2019}. Promising approaches to increase the coding performance for such applications are, e.g., based on saliency coding as in~\cite{choi2018} and~\cite{galteri2018}. Another approach in~\cite{fischer2020_ICIP}, optimizes the in-loop filtering of VVC for the VCM task. The same authors modified the rate-distortion optimization inside the VVC encoder to improve the coding performance when Mask R-CNN is applied to the decoded frame~\cite{fischer2020_FRDO}.

Moreover, several approaches have already been made to improve the robustness of classification networks. The authors in~\cite{dodge2016} investigated the classification accuracy of neural networks for input data that has been affected with blurriness, noise, and Joint Photographic Experts Group~(JPEG) compression. In~\cite{borkar2019_IEEE}, the convolution layers of a classification network that performed worse on distorted input data according to a self-defined metric were retrained to improve the detection accuracy for distorted images. Another approach in ~\cite{ghosh2018} proposed adapting weights according to the incoming image degradation by a master-slave architecture.

\section{Background R-CNNs}

Faster R-CNN~\cite{ren2017} is a deep convolutional neural network detecting objects from an input image by first applying convolutional layers to transform the input image into the feature space. Subsequently, a region proposal network (RPN) is applied to this feature space in order to propose possible objects. Each of the resulting regions of interest (RoIs) in the feature space is fed into a so-called region of interest pooling layer that returns an output feature map of fixed dimensions. Thus, this feature map can eventually be fed into a fully-connected network called classifier that computes for each proposal which object class the proposed object belongs to. There, it is also possible that the RoI is classified as background and is thus not considered as relevant object anymore. In addition, the bounding box is refined.

During training, the RPN and the other parts, which include the convolutional layers and the classifier, are trained iteratively~\cite{ren2017}. First, the convolutional layers are pre-initialized with weights from an image classification model. Then, these weights are frozen and the RPN is trained by reducing the loss $\mathcal{L}_{\mathrm{RPN}}$, which is derived from the classification loss $\mathcal{L}_{\mathrm{cls}}$ and the regression loss $\mathcal{L}_{\mathrm{reg}}$ by

\begin{equation}
\mathcal{L}_{\mathrm{RPN}} = \mathcal{L}_{\mathrm{cls}} + \lambda \cdot \mathcal{L}_{\mathrm{reg}}.
\end{equation}
Thereby, the classification loss $\mathcal{L}_{\mathrm{cls}}$ is calculated by the binary cross-entropy defining whether an RoI has been correctly identified and the loss $\mathcal{L}_{\mathrm{reg}}$, which calculates how well the bounding box of the RoI fits to the ground-truth object. The second loss is only considered when there actually is a ground-truth label at a certain position. $\lambda$ is used as a weighting factor and is set to 10 in the original paper~\cite{ren2017}.

Afterwards, the convolutional layers to obtain the feature space and the classifier are trained with fixed RPN weights from the previous step. Subsequently, the RPN is trained again with the new weights. In the following, this process is repeated iteratively until the overall loss is minimized. For a detailed explanation of the whole training process, the reader is referred to~\cite{ren2017} and~\cite{girshick2015}.

The structure of Mask R-CNN~\cite{he2017} is identical, besides an additional branch consisting of convolutional layers which returns a binary mask for each detected object to segment it pixel-wisely from the background.

\section{Proposed Training Procedure with Compressed Data}

Classic training frameworks of object detection and segmentation networks like Faster and Mask R-CNN train the weights and biases of the models with a set of pristine and uncompressed input images $\boldsymbol{X}$. For this, ground truth data is required, which consists of the objects $\boldsymbol{P}$ including a class label and a bounding box or binary mask to detect. With this data, the loss function can be minimized until the weights are adapted to the given task. However, in a practical application as shown in Fig.~\ref{fig:edge computing scenario}, the input data $\boldsymbol{\tilde{X}}$ suffers from compression artifacts reducing the detection accuracy of the network significantly, especially when only low data rate is available. Derived from this, we present two different training possibilities to increase the R-CNN performance when being applied to such deteriorated data in the following.

\subsection{Data Augmentation}

Data augmentation is a widely used process to improve the training~\cite{shorten2019}. E.g. for classification networks, the input images are additionally rotated or flipped and fed into the network, which has two major advantages. First, the network learns a more general concept of the objects that have to be classified and is thus more robust to changes during inference. Second, the number of required images inside a training dataset is smaller, which helps when only a small dataset is available.

As proposed data augmentation, we compressed each image $X_i$ with index $i$ from the pristine input dataset before training and added it to the training data. All in all, the final dataset thus includes the uncompressed input data $\boldsymbol{X}$ and the compressed images $\boldsymbol{\tilde{X}}_c$ for a certain number of different compression levels $c$. As compression method, we select the same codec, VVC test model~(VTM)~\cite{chen2020vtm10} in all intra configuration, as it is also applied in the inference scenario investigated in Section~\ref{sec:Results}. The compression parameter $c$ is called quantization parameter~(QP) in VTM and was set to values of 22, 27, 32, 37, 42, and 47, with $\boldsymbol{X}$ and each $\boldsymbol{\tilde{X}}_c$ containing the same amount of images. With this, one model can be obtained that covers the possible bitrates during inference without having to train one model for each possible bitrate point.

We additionally trained another model in the same manner, but replaced the VTM compression with the image codec JPEG2000~\cite{christopoulos2000} and its reference software~\cite{openjpeg2000} to further evaluate whether the proposed training method also works when the compression scheme used during inference is not known before training. To that end, compression parameters corresponding to the resulting PSNR of 33, 35, 37, 40, 45, and 50 were chosen.

\subsection{Fine-Tuning}

For fine-tuning, contrary to data augmentation, the network is first trained with the pristine, uncompressed dataset for a certain number of iterations. Afterwards, the training data is replaced and the weights are fine tuned with the compressed dataset for an additional number of iterations. We used this procedure to first train the networks with the pristine data $\boldsymbol{X}$, as in the classic training, and then used the compressed images $\boldsymbol{\tilde{X}}_c$ for making the weights more robust against image degradations induced by compression during inference.

\begin{table}[t]
	\centering
	
	\caption{Number of iterations used for training.}
	\small
	\def\rowWidth{1.8cm}
	\begin{tabular}{l|p{\rowWidth}p{\rowWidth}}
		
		Training Method & Faster \newline R-CNN & Mask \newline R-CNN \\
		\hline
		\rule{0pt}{1.2\normalbaselineskip}Classic training & 25,000 & 24,000\\
		Data augmentation & 35,000 & 34,000 \\
		Fine-tuning & 10,000 & 10,000 \\
		
	\end{tabular}
	\label{tab:parameters data augmentation}
	\vspace{\VSPACE}
\end{table}

\subsection{Training Parameters}

As training framework for Faster and Mask R-CNN, we used the \textit{Detectron2} library~\cite{wu2019detectron2} including \textit{PyTorch} implementations for each R-CNN. For both networks, the same ResNet50~\cite{he2016resnet} backbone with feature pyramid structure was taken. The models were initialized with weights from~\cite{wu2019detectron2} trained on the COCO-dataset~\cite{lin2014_COCO} before training. We mainly stayed with the default training hyper parameters from the Detectron2 library. For Faster R-CNN, a learning rate of 0.00025 and a batch size of 7 images was selected. For Mask R-CNN, the learning rate was initially set to 0.01 and decreased to 0.001 after 18,000 iterations with a batch size of 2 images similar to~\cite{he2017}. The defined numbers of iterations are listed in Table \ref{tab:parameters data augmentation} for both R-CNNs. For fine-tuning, the model trained with the pristine dataset has been taken and fine-tuned for additional 10,000 iterations in order to have a comparable number of iterations between data augmentation and fine-tuning.

\section{Experimental Setup}

\subsection{Dataset}
\label{subsec:Dataset}

A suitable dataset is required for our experiments, which combines two aspects. First, the dataset has to be stored as uncompressed image data and, second, requires labeled data. Thus, the \textit{Cityscapes}~\cite{cordts2016} dataset was taken, which includes pixel-wisely labeled data of eight different types of road users captured with a stereo camera mounted behind the front shield of a car. For training, the 2975 images from the training dataset were considered.

\subsection{Evaluation Setup}

In order to evaluate the performance of the proposed robust models on compressed data, the 500 images with a size of $2048\times1024$ pixels from the \textit{Cityscapes} validation dataset were coded with the standard-compliant VTM-10.0~\cite{chen2020vtm10} and QP values of 22, 27, 32, 37, 42, and 47 in all-intra configuration similar to~\cite{bossen2019}. The detection accuracy of Faster and Mask R-CNN was evaluated with the average precision (AP) for each class in the version presented for \textit{Cityscapes}~\cite{cordts2017cityscapesscripts}. Besides, the APs over the different classes were weighted according to their appearance in the validation dataset as proposed in~\cite{fischer2020_ICIP} and~\cite{fischer2020_FRDO}. This metric is called weighted AP in the remainder of this paper.

\section{Results}
\label{sec:Results}

\subsection{Weighted AP over Bitrate}


\begin{figure}[t]{}
	\centering
	\includegraphics[width=0.5\textwidth]{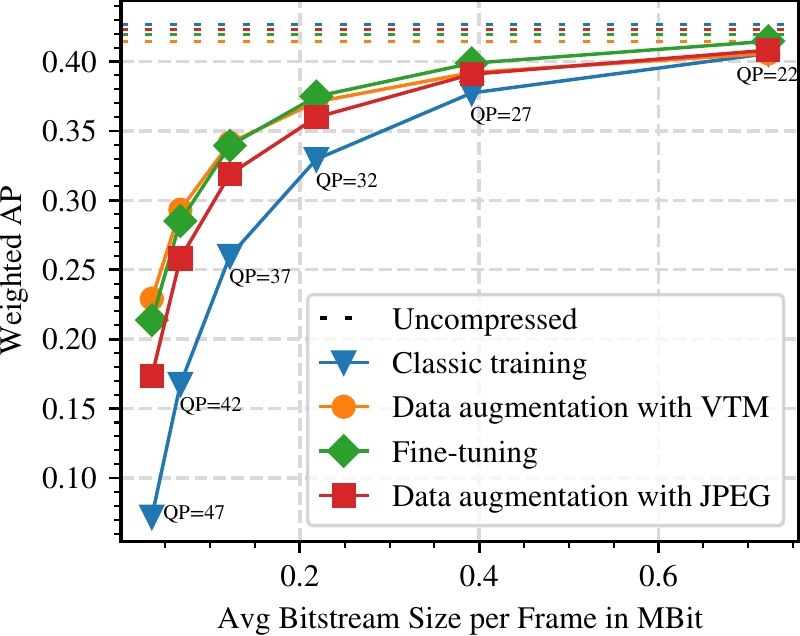}
	\caption{Weighted AP values over bitrate comparing the different training methods for applying Mask R-CNN to the decoded images. The dotted lines represent the weighted AP on uncompressed input images for the respective methods.}
	\label{fig:weighted ap over bitrate mask r-cnn}
\end{figure}

Fig.~\ref{fig:weighted ap over bitrate mask r-cnn} plots the weighted AP when applying the different Mask R-CNN models to the input data compressed with the VTM for different QPs. The abscissa shows the bitrates. When applying the models to the uncompressed and pristine input data (dotted lines), the model trained with classic training performs best with a weighted AP of 0.43. With decreasing bitrate, all models perform worse due to the diminishing information and the increasing compression artifacts. However, the models that have been trained with the proposed training methods of data augmentation and fine-tuning perform significantly better than the model that has exclusively been trained with pristine data. Fine-tuning works slightly better than data augmentation for higher bitrates, whereas the Mask R-CNN model trained with data augmentation performs marginally better for the highest investigated QPs of 42 and 47.

The model that has been trained with JPEG2000 compressed input data for data augmentation also outperforms classic training, but does not reach the coding gains that are achieved when training with VTM compressed data. This shows that the models can generally be trained to the aspect of quantization and distortions induced by lossy compression. However, to achieve the highest model robustness for the input data to be expected, it is inevitable to perform training with the data covering unique characteristics induced by, e.g., block-partitioning or in-loop filters of the target codec.

%
%
%

%
%


\subsection{Bj\o ntegaard Delta Values}

In order to quantify the coding gains of the curves shown in Fig.~\ref{fig:weighted ap over bitrate mask r-cnn}, we calculated the Bj\o ntegaard delta values~\cite{bjontegaard2001_new} for the QPs suggested by JVET~\cite{bossen2019} and replaced the commonly used PSNR with the weighted AP metric as it is also suggested by the JVET VCM group~\cite{liu2020_VCM_CTC}. First, Table~\ref{tab:BD weighted AP values} lists the gains in weighted AP for the same bitrate when one of the proposed training methods is taken over the classic training. For both R-CNNs, fine-tuning leads to the highest weighted AP gains of up to 3.68 percentage points. Furthermore, the proposed methods achieve similar coding gains for both considered R-CNNs.

From another perspective, the Bj\o ntegaard delta rate values in Table~\ref{tab:BD rate values} denote how much bitrate can be saved at the same detection accuracy, when using the proposed training. For the best case, almost 48\,\% of bitrate can be saved when using the Faster R-CNN Model trained with fine-tuning. We additionally evaluated the coding gains for very low bitrate, the four highest QPs, which results in even higher coding gains of up to 60\,\%. However, these low bitrate areas are not often targeted in practice, since a high accuracy near to the accuracy on uncompressed data is vital for security-relevant real-life applications.

\begin{table}[t]
	\centering
	\small
	\caption{Bj\o ntegaard delta weighted AP values in percentage points of weighted AP with the classic training method as anchor for QP of 22, 27, 32, and 37.}
	\def\rowWidth{1.5cm}
	\begin{tabular}{l|p{\rowWidth}p{\rowWidth}}
		
		Training Method & Faster \newline R-CNN & Mask\newline R-CNN \\
		\hline
		\rule{0pt}{1.2\normalbaselineskip}Data augmentation VTM & 3.07 & 3.08\\
		Fine-tuning & 3.68 & 3.57 \\
		Data augmentation JPEG &  2.01 & 2.38 \\
		
	\end{tabular}
	\label{tab:BD weighted AP values}
	\vspace{\VSPACE}
\end{table}

\begin{table}[t]
	\centering
	\small
	\caption{Bj\o ntegaard delta rate values in percentage with the classic training method as anchor for QP of 22, 27, 32, and 37. Negative values represent bitrate savings compared to the anchor.}
	\def\rowWidth{1.5cm}
	\begin{tabular}{l|p{\rowWidth}p{\rowWidth}}
		
		Training Method & Faster \newline R-CNN & Mask\newline R-CNN \\
		\hline
		\rule{0pt}{1.2\normalbaselineskip}Data augmentation VTM & -40.53 & -35.59\\
		Fine-tuning & -47.98 & -42.59 \\
		Data augmentation JPEG & -26.77 & -28.31\\
	\end{tabular}
	\label{tab:BD rate values}
	\vspace{\VSPACE}
\end{table}

\subsection{Visual Example}

In Fig.~\ref{fig:Visual Comparison}, a visual comparison is provided between applying the classically trained Mask R-CNN model and the model trained with the proposed data augmentation to an exemplary image that has been compressed with a QP of 42. The model that has only been trained with pristine data is not able to detect small objects from the heavily quantized input data. Contrary, the model that has been trained with the proposed data augmentation is still able to detect those objects, which might be vital for subsequent processing such as hazard detection.


\def\subwidth{0.49}
\begin{figure}[t]
	\centering
	\begin{subfigure}[t]{\subwidth\textwidth}
		\centering
		\includegraphics[width=\textwidth]{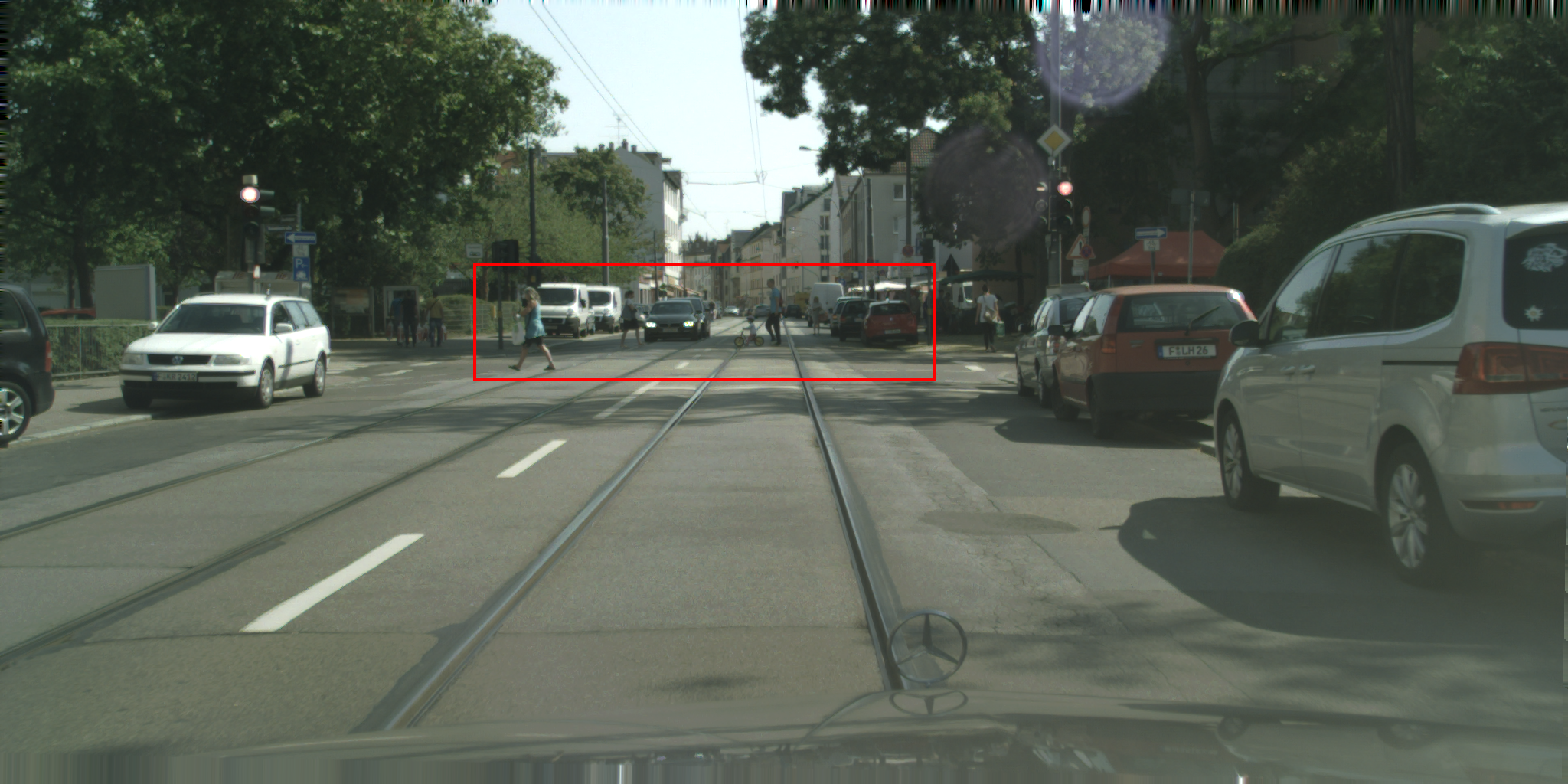}
		\caption{Uncompressed image} \label{fig:comp1}
	\end{subfigure}
	\begin{subfigure}[t]{\subwidth\textwidth}
		\centering
		\includegraphics[width=\textwidth]{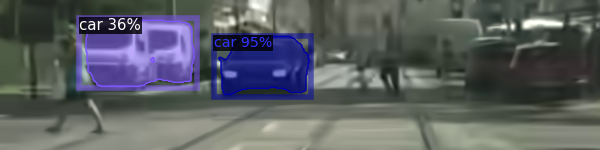}
		\caption{Classic training} \label{fig:comp2}
	\end{subfigure}
	\begin{subfigure}[t]{\subwidth\textwidth}
		\centering
		\includegraphics[width=\textwidth]{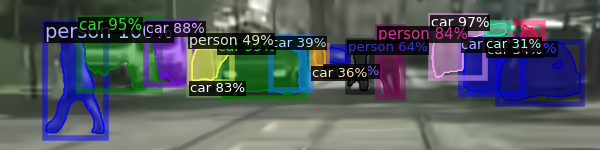}
		\caption{Proposed data augmentation} \label{fig:comp3}
	\end{subfigure}
	\vspace{2mm}
	\caption{Visual comparison applying the Mask R-CNN model with classic training and the model with proposed data augmentation to \textit{Cityscapes} image \textit{frankfurt\_000000\_001236} compressed with VTM and a QP of 42. The size of the shown excerpt marked with the red frame in (a) is $600\times150$ pixels}
	\label{fig:Visual Comparison}
\end{figure}

\vspace{\VSPACE}
\section{Conclusions}

In this paper, we propose to train the object detection and segmentation networks Faster and Mask R-CNN additionally with compressed data in order to make them more robust against deteriorated input data, which occurs in real world applications. With specific fine-tuning, the weighted average precision can be increased by up to 3.68 percentage points compared to a model that is exclusively trained with pristine data. When applying this model in real-world scenarios up to 48\,\% bitrate could be saved while still obtaining the same detection accuracy. In future work, it could also be investigated whether training and applying the models specifically to only one QP delivers even higher coding gains. However, in practical scenarios, it might be more feasible to only apply one jointly trained model for all bitrates. In addition, the evaluation might also be conducted for video data. Then, the robustness could be further increased by adding additional side-information derived from the decoding process like motion information.

\newpage
\bibliographystyle{IEEEtran}
\bibliography{/home/fischer/Paper/jabref_literature_research_ms2.bib,/home/fischer/Paper/literature_M2M_communication.bib,/home/fischer/Paper/jabref_used_software.bib}

\end{document}